\documentstyle[preprint,tighten,eqsecnum,epsf,aps]{revtex}

\def\dj{d\kern-.30em\raise1.25ex\vbox{\hrule width .3em height .03em}} 
\def\Dj{D\kern-.75em\raise0.75ex\vbox{\hrule width .3em height .03em}}

\def\arctan{\mathop{\rm arctan}\nolimits}

\begin{document}

\draft

\title{The Effect of Low Momentum Quantum Fluctuations on a Coherent Field 
Structure}
\author{G. Cruz-Pacheco, A. Minzoni, P. Padilla}
\address{FENOMEC-IIMAS\\
Universidad Nacional Aut\'onoma de M\'exico}
\author{ 
A. Corichi, M. Rosenbaum, M. P. Ryan}
\address{
Instituto de Ciencias Nucleares - FENOMEC\\
Universidad Nacional Aut\'onoma de M\'exico\\
A. Postal 70-543, M\'exico D.F. 04510, M\'exico.}
\author{
N.F. Smyth}
\address{
Department of Mathematics and Statistics, University of Edinburgh,\\
The King's Buildings, Mayfield Road,\\
Edinburgh, Scotland, U.K., EH9 3JZ.}
\maketitle

\begin{abstract}
In the present work the evolution of a coherent field structure of the 
Sine-Gordon equation under quantum fluctuations 
is studied.  The basic equations are derived from the coherent state 
approximation to the functional Schr\"odinger equation for the field.  These 
equations are solved asymptotically and numerically for three physical
situations.  The first is the study of the nonlinear mechanism
responsible for the quantum stability of the soliton in the
presence of low momentum fluctuations.  The second considers the
scattering of a wave  by the Soliton.  
Finally the third problem considered is  
the collision of Solitons and the stability of a breather.
  It is shown that 
the complete integrability of the Sine-Gordon equation precludes fusion and
splitting processes in this simplified model.  
  The approximate results obtained are non-perturbative in 
nature, and are valid for the full nonlinear interaction in the limit of low 
momentum fluctuations.  It is also found that these approximate results are 
in good agreement with full numerical solutions of the governing equations.
This suggests that a similar approach could be used for
the baby Skyrme model, which is not completely integrable.
In this case the higher
space dimensionality and the internal degrees of freedom which prevent
the integrability will be responsable for fusion and splitting 
processes. This work provides a starting point in the numerical solution
of the full quantum problem of the interaction of the
field with a fluctuation.
\end{abstract}


\section{INTRODUCTION}

In the past few years it has been shown that the Skyrme model \cite{skyrme}
is related to the low energy limit of QCD \cite{witten:83}.  This fact,
together with progress in the approximate and numerical solution
of strongly nonlinear equations, has renewed interest in a detailed 
study of the quantum mechanics of the Skyrme model. These studies are directed 
along two main lines.  The first simplifies the model at the classical level 
to the so-called Baby Skyrmion model, which in turn is related to the 
Sine-Gordon equation \cite{skyrme,KPZ,PZ}.  In this simplified classical 
model extensive numerical studies have led to an increased understanding 
of the stability, scattering and interaction of coherent structures with 
radiation \cite{KPZ,PZ}.  On the other hand, the second line has
focused on the quantum effects of the full Skyrme model
\cite{adkins,gisiger}. In 
these studies the quantization was either obtained by linearization around 
a field configuration \cite{adkins} or by a finite dimensional 
approximation to the full problem \cite{gisiger}.

Here we take a complementary approach. We study the 
1+1 Sine-Gordon equation keeping all the degrees of freedom and quantize along the lines of \cite{eboli1,eboli2,ours,cooper}. This leads to a field equation which is   strongly coupled with the equations for the fluctuations. In \cite{cooper} the formalism for the functional coherent state approximation was fully developed and the possible advantages and disadvantages of various approaches designed for numerical purposes were discussed in detail. In particular, the closed-time path method with the Hartree factorization (see e.g. \cite{keldish}) was applied in \cite{cooper} to the static Sine-Gordon field, and the approximate results obtained for phase transitions and stability were found to compare favorably with known exact results. In this same work static results were also obtained for more realistic field equations.
Our present work differs from the above cited papers in that we choose to approximate the Green function (the variance kernel for the Gaussian ansatz) by a suitable parametrized trial function. This choice leads to a great simplification of the problem for the case of low momentum fluctuations. Thus, for example, 
the infinite system of partial differential
equations of \cite{eboli1} become ordinary differential equations and, eventhough the field and the fluctuations in our formalism are strongly 
coupled, we are able to use some of the solutions of the classical Sine-Gordon equation (since in $1+1$ dimensions it is completely integrable) to construct 
approximate solutions to the quantum problem, including the effect of the
radiation generated by the quantum fluctuations. We then solve the equations  
numerically and these numerical solutions are compared with asymptotic 
solutions.
Note also that the approach used in the present 
work is completely different to that of  \cite{lu,lu2}  
where the wave functionals are constructed using
Gaussian approximations to the functional 
Schr\"odinger equation for the Sine-Gordon field.  
However, the particles are considered as higher excited
states (in function space) of the linearized field 
equations.  In our treatment the field equations are nonlinear and dynamic 
and different particles are represented by different nonlinear 
field configurations,
not by higher order Hermite functionals as in \cite{lu,lu2}.

Finally, it must be stressed that the approximate analytic results 
obtained here are valid in a strongly nonlinear regime and, in 
principle, do not depend on the $1+1$ nature of the model and 
could be used to study problems related to more realistic 
simplifications of the Skyrme model.

The paper is organized as follows.  
In Section II the detailed formulation of 
the quantum problem is stated, with the free parameters adjusted
to mimic mesons and baryons.  
Section III is devoted to the study of the
coherent state approximation and the derivation of the quantum equations
for the field for low momentum fluctuations.  
In Section IV three problems are considered.  
The first is the nonlinear stability of a single Soliton under 
the influence of quantum fluctuations.  This stability is studied both 
numerically and asymptotically.  In particular the asymptotic
solution includes
the damping effect of the radiation shed by the Soliton due to the 
fluctuations.  This asymptotic solution explains in detail the
mechanism for the nonlinear stability of the Soliton. 
 The second problem studied is the scattering 
of a meson (wave) by a static Soliton.  The numerical solution for this 
problem shows that in this process the Soliton is also stable. 
 Finally the 
third problem studied concerns the collision of Solitons and the quantum 
evolution of a bound state (Soliton and an anti-Soliton). 
 It is shown that 
the complete integrability of the Sine-Gordon
simplification of the full Skyrme model to 
just one internal degree of freedom for the 
field precludes the processes of 
fusion and splitting.  
The processes of fusion and splitting are produced by 
the influence of all the internal degrees of freedom.

\section {FORMULATION OF THE PROBLEM}

For the basic structure we take the baby Skyrme model, which is a reduction
of the full model with only two fields present \cite{skyrme,KPZ}.
This model, in turn, reduces to the Sine-Gordon equation which is known
to be completely integrable \cite{KPZ}.
In these variables the Hamiltonian takes the form
\begin{equation}
H = m {c}^2 l \int {\left[ \frac{1}{2} \pi^2 + \frac{1}{2}
\left( \frac{\partial\varphi}{\partial x} \right)^2 + l^{-2} 
(1 - \cos \varphi)\right] dx},
\label{e:ham}
\end{equation}
where $\varphi$ is an angle variable whose momentum $\pi$ is given by
\begin{equation}
\pi =\frac{1}{c} \frac{\partial\varphi}{\partial t},
\label{e:pi}
\end{equation}
$m$ is the mass of the particle and $l$ is a typical particle size.
Using dimensionless variables $\tilde x = x/l $ and $ \tilde t = ct/l$, we obtain, after dropping the tildes,
\begin{equation}
H = m {c}^2 \int {\left[ \frac{1}{2} \pi^2 + \frac{1}{2}
\left( \frac{\partial\varphi}{\partial x} \right) ^2 + (1 - 
\cos \varphi)\right] \: dx},
\label{e:nondimham}
\end{equation}
with $\pi = \partial\varphi / \partial t$.  
To mimic a baryon by means of 
the Sine-Gordon soliton, we take $l \sim 10^{-13}$ cm 
and $ m \sim 10^{-27}$ Kg, 
which gives an internal time of $10^{-23}\rm {sec}$.

The equation of motion derived 
from the Hamiltonian variational principle
\begin{equation}
\delta \int_{t_0}^{t_1} \left( \pi \dot{\varphi} - H \right) \: dt
\label{e:varham}
\end{equation}
is the Sine-Gordon equation
\begin{equation}
\frac{\partial^2 \varphi}{\partial t^2} -
\frac{\partial^2 \varphi}{\partial x^2} + \sin \varphi =0.
\label{e:sineg}
\end{equation}
The one dimensional Skyrmion is the soliton solution
\begin{equation}
\varphi = - 4 \arctan \left[  \exp \left( - (x - vt) / \sqrt{1-v^2} 
\right) \right],
\label{e:sgsol}
\end{equation}
of \ref{e:sineg}, which represents a localized deformation at $x = vt$.  In 
this context $v$ must be taken to satisfy $v \ll 1$, since the Skyrme model 
is only consistent for small energies.  The linear travelling periodic wave 
solutions of the Sine-Gordon equation are interpreted as pions.
Notice that the nonlinear model contains linear waves which describe
bosons and nonlinear structures which describe fermions.  
In his original work
Skyrme suggested that the fermionic part of the Lagrangian is needed only just
to count the number of localized states of finite amplitude of the bosonic
field \cite{skyrme}.  In this interpretation, fermions are just a point
approximation to nonlinear localized bose fields.  It has also been
suggested on rather general grounds that the canonical quantization of fields
(as bosons) gives Fermi-Dirac type statistics for the kinks \cite{finkelstein}.
In the Sine-Gordon model the exclusion principle 
holds for kinks, since we
know that for the general exact solution there is no solution with two
identical solitons \cite{lamb}.

In this article, we shall consider a canonical quantization using the
functional Schr\"odinger picture.  It is to be noted that all quantizations 
for the Skyrme model cited in the Introduction make the same assumption. 
However, in this work we differ from previous treatments in that we shall keep
infinitely many degrees of freedom in the classical field and reduce the 
dimensionality of the space of fluctuations.  The final result will be shown 
to be a system which consists of a partial differential equation (similar to 
the Sine-Gordon equation) for the field coupled to a system of nonlinear 
ordinary differential equations which control the fluctuations.

In the field configuration representation the 
Schr\"odinger equation takes the form
\begin{equation}
i \hbar \frac{\partial \Psi(\varphi)}{\partial t} = \hat H \Psi(\varphi),
\label{e:5}
\end{equation}
where $\Psi(\varphi(x), t):= \langle \varphi(x) |\Psi(t)\rangle$ is the
amplitude for finding the field system characterized by the state vector
$|\Psi(t)\rangle$ in the field configuration $\varphi(x)$ at time $t$.
In this configuration representation the scalar product of two state
vectors is given by the functional integration
\begin{equation}
\langle \Psi_{1} | \Psi_{2} \rangle = \int {\Psi^\ast _{1}(\varphi,t)}
\Psi_2 (\varphi,t) \: {\cal D} \varphi, 
\label{e:funint}
\end{equation}
and the field operators are represented by functional kernels.  Thus the
field operator $\hat \phi(x)$ is represented by $\langle \varphi(x) |
\hat \phi(x)|\Psi(t)\rangle = \varphi(x) \Psi(\varphi(x),t)$ and, therefore,
acts as a multiplication operator, while the action of the canonical
momentum operator is given by 
\begin{equation}
\langle \varphi(x)|\hat \pi (x')|
\Psi(t)\rangle = -i \hbar \frac{\delta}{\delta \varphi(x')} \langle
\varphi(x) |\Psi(t)\rangle . 
\label{e:conmom}
\end{equation}
The Hamiltonian field operator $\hat H$ then becomes
\begin{equation}
\hat H = mc{^2} \int \left[ \frac{1}{2} {\hat \pi}^2 (x) + \frac{1}{2}
\left( \frac{\partial\hat \varphi}{\partial x} \right)^{2} + 
(1 - \cos \hat \varphi) \right] \: dx,
\label{e:fieldop}
\end{equation}
where there is no ambiguity in the ordering and 
the functions of operators
are defined by their power series.

The quantum mechanical problem for the field consists of solving the
Schr\"odinger equation (\ref{e:5}) for a given initial field configuration. 
Notice that the time in (\ref{e:5}) has a scale set by $mc^{2} / \hbar$, 
which is of the same order of magnitude ($ 10^{-23}$) as the time scale 
$c / l$ for the field fluctuations.  This is to be expected since the Skyrme 
equation was assumed to be consistent with the quantum scale of the particle. 
Thus we can take the same dimensionless time variable for either scale.

\section{APPROXIMATE SOLUTIONS TO THE FUNCTIONAL EQUATION}

To solve the Schr\"odinger equation (\ref{e:5}), we take a coherent state 
approximation
\cite{eboli1,eboli2,ours,cooper} and study the evolution of its parameters.  
Following  \cite{eboli1,eboli2,ours,cooper} we consider the functional
\begin{equation}
\Gamma = \int \langle \Psi | i \frac{\partial}{\partial t} - 
\hat H | \Psi
\rangle \: dt,
\label{e:gamma}
\end{equation}
which we extremize with the Gaussian trial functional
\begin{eqnarray}
\Psi(\varphi(x) ,t)&=& \exp \left\{ i \int \pi (x,t) \left[ \varphi (x) -
\phi (x,t)\right] \: dx \right\}  \nonumber \\
 & & \: \exp \left\{ - \int \int dx dy \: \left[
\varphi (x) - \phi (x,t) \right] \right. \nonumber \\
          & & \left. (\times)\left[ \frac{1}{4} \Omega ^{-1} (x,y ,t) - i
\Sigma (x,y,t) \right] \left[ \varphi (y) - \varphi (x) \right] \right\} ,
\label{e:trial}
\end{eqnarray}
where the kernels $\Omega^{-1}$ and $\Sigma$ take into account the quantum
fluctuations and $\phi(x,t)$ and $\pi (x,t)$ are the average field and
average momentum of the Gaussian, respectively.

Substituting the trial function (\ref{e:trial}) into the functional
(\ref{e:gamma}) and integrating over the $\varphi(x)$ variable
yields an averaged action \cite{eboli1,eboli2,ours,cooper}.
This action is given in terms of
$\phi(x,t)$, $\Omega$ and $\Sigma$.  The potential is then also expanded 
around the average $\phi(x,t)$.  It should be stressed that the consistency 
of this approximation depends on the smallness at all times of the variance 
$\Omega$, in the sense that the average energy of the fluctuations around 
the mean $\phi(x,t)$ is small compared with the energy of the average motion. 
Moreover, by keeping only quadratic terms resulting from the averaging around 
the mean of the nonlinear term, which amounts to making the assumption that 
the energy of the fluctuations is small compared to the energy of the mean, 
we arrive at an effective action of the form
\begin{eqnarray}
L &=& \int_{0}^{T} \left\{ \int_{-L}^{L} \left[ \pi \frac{\partial
\varphi}{\partial t} - \left( \frac{1}{2} \left(\frac{\partial \varphi}
{\partial x}\right)^{2} + (1 - \cos \varphi )\right)\right. \right. \nonumber\\
  & & \mbox{} + \Sigma {\bf\dot{\Omega}}(x,x,t) -2 \Sigma \Omega \Sigma 
(x,x,t) - \cos (\varphi) \Omega(x,x,t)\nonumber \\
  & & \mbox{} \left. \left. + \frac{1}{2} \frac{\partial^{2} \Omega}{\partial 
x^{2}}(x,y,t)|_{x=y} - \frac{1}{8}\Omega^{-1}(x,x,t) \right] \right\} \: 
dxdt + O(\Omega^{2}),
\label{e:effaction}
\end{eqnarray}
where the notation $\Omega \Sigma$ is taken to mean the kernel of the
operator defined by the convolution of $\Sigma$ with $\Omega$, that is
\begin{equation}
\Omega \Sigma (x,y) = \int_{-L}^{L} \Omega(x,z) \Sigma(z,y) \: dz
\label{e:conv}
\end{equation}
and $\Sigma \Omega \Sigma$ is the kernel of the operator defined by the
convolution of $\Sigma$, $\Omega$ and $\Sigma$
\begin{equation}
\Sigma \Omega \Sigma (x,y) = \int_{-L}^{L} \int_{-L}^{L}
\Sigma (x,z) \Omega (z,u) \Sigma (u,y) \: du dz.
\label{e:sos}
\end{equation}
In the following we will take the limit $L \rightarrow \infty$ at different stages  in the calculation of the effective action.
The effective action can be computed once we choose an appropriate
parameterization for the variance $\Omega$ and the phase $\Sigma$.
It must be noted that thanks to the simple form of the potential.
the Gaussian integral can be evaluated exactly. However, since
we are interested in small fluctuations we stop at the quadratic
level. We will come back to this point when comparing our results
with those in \cite{cooper}. 

Now, since the field is homogeneous, we can take
\begin{eqnarray}
\Omega (x,y,t)&=& \frac{1}{2 \pi} \int_{-\infty}^{\infty} \exp [ik(x-y)] \hat
\Omega (k,t) \: dk
\label{e:hom1} \\
\Sigma (x,y,t)&=& \frac{1}{2\pi}\int_{-\infty}^{\infty}
\exp [ik(x-y)] \hat \Sigma (k,t) \: dk, 
\label{e:hom2}
\end{eqnarray}
where
\begin{eqnarray}
\hat \Omega (k,t)&=& \frac{\Omega_{0}}{k^{2} + a^{2}(t)}, \label{e:omhat} \\
\hat \Sigma (k,t)&=& \frac{b(t)}{k^{2} + a^{2}(t)}. \label{e:sighat}
\end{eqnarray}

The parameters $a(t)$ and $b(t)$, which control the spreading of the
fluctuations, are to be determined after the effective action is varied.  
The choice of the trial function is then guided by the simplicity of the 
resulting expressions.  Notice, however, that since the results obtained 
depend only on the spreading, the same qualitative behaviour will be obtained 
for other forms of the trial function.  

Since the approximate solution (\ref{e:trial}) involves the kernel 
$\Omega ^{-1}$ the proposed expression is convergent provided that the 
momentum of the fluctuations involved in the integration is low.  This 
assumption is in agreement with the fact that the basic Skyrme model is 
consistent at low momentum and the assumed homogeneity of the fluctuation.  This is taken into account by taking for 
$\Omega ^{-1}(x,y,t)$ the cut-off kernel 
\begin{equation}
\Omega ^{-1} (x,y,t) = \frac{1}{2\pi \Omega _{0}} \int _{-K}^{K} e^{ik(x-y)} \: 
\left( k^{2} + a^{2} \right) \: dk.
\label{e:kernal}
\end{equation}
Then the non-constant contribution of $\Omega ^{-1}(x,x,t)$ is
\begin{equation}
-\frac{1}{8} \Omega ^{-1} (x,x,t) = -\frac{K}{8\pi \Omega_{0}} a^{2}.
\label{e:contrib}
\end{equation}
In a similar manner we obtain
\begin{eqnarray}
\frac{1}{2} \frac{\partial ^{2}}{\partial x^{2}} \Omega (x,y,t) |_{y=x} 
& = & \lim _{K \to \infty} -\frac{1}{4\pi} \Omega_{0} \int _{-K}^{K} \frac{k^{2}}
{k^{2}+a^{2}} \: dk \nonumber \\
 & = & \lim _{K \to \infty} \left[ -\frac{\Omega_{0} K}{2\pi} + \frac{1}{4\pi} a(t) \Omega_{0} 
\int _{-\infty}^{\infty} \frac{dk}{k^{2}+1} \right] .
\label{e:limit2}
\end{eqnarray}
The first term in this expression is infinite, but a constant.  Thus it does 
not contribute to the equations of motion.  We therefore take just the second
term in the effective Lagrangian.  Finally the parameter $\Omega_0$ 
measures the size of the fluctuations.  In principle, other choices of the 
parameters may lead to different regularizations.  However, as discussed below,
the basic qualitative picture described in this work is not changed by these 
alternative regularizations, provided the momentum is low. 
For the case of higher momentum, procedures similar to the 
ones discussed in \cite{eboli2,cooper}
 can be used.

With the above assumptions, the effective Lagrangian takes the form
\begin{eqnarray}
L &=& \int_{0}^{T} \int_{-L}^{L} \left\{ \pi \frac{\partial
\varphi}{\partial t} - \left( \frac{1}{2} \left(
\frac{\partial \varphi}{\partial
 x}\right)^{2} + \left[ 1 - \left( 1-\frac{\Omega_{0}}{2\pi a} c_{2}\right) 
\cos \varphi \right] \right) \right. \nonumber\\
  & & \mbox{} \left. + \frac{1}{2\pi} \left( - 2 \frac{\dot{a} b}{a^{4}}\Omega_{0} c_{1} 
-2\frac{b^{2}}{a^{5}} \Omega_{0} c_{1} - \frac{K}{4\Omega_{0}}
a^{2} + \frac{1}{2} \Omega_{0} c_{2} a \right)\right\} \: dx dt,
\label{e:efflang}
\end{eqnarray}
where the constants $c_{1}$ and $c_{2}$ are given by
\begin{equation}
c_{1}= \int_{-\infty}^{\infty} \frac{dk}{(1+k^{2})^{3}}=\frac{\pi}{2},  
\qquad c_{2}= \int_{-\infty}^{\infty} \frac{dk}{(1+k^{2})}= \pi.
\label{e:const}
\end{equation}

It is now convenient to change variables and define $q=\frac{1}{a^{3}}$ and
$b=p$ in order to obtain the Lagrangian (\ref{e:efflang}) in the form
\begin{eqnarray}
L &=& \int_{0}^{T} \left\{ \left( \frac{2 c_{1} \Omega_{0}}{3} \dot{q} 
p - 2c_{1}\Omega_{0} p^{2} q^{-\frac{5}{3}} - \frac{K}{4\Omega_{0}}
q^{-2/3} + \frac{1}{2} \Omega_{0} \frac{c_{2}}{c_{1}} q^{-1/3} \right) \: \frac{L}{\pi} dt
\right. \nonumber \\
  & & \mbox{} + \left. \int_{-L}^{L} \left[ \pi \frac{\partial \varphi}
{\partial t}- \frac{1}{2} \pi^{2} - \frac{1}{2} \left( 
\frac{\partial \varphi} {\partial x}\right)^{2} - 
[1-(1- \frac{\Omega_{0}}{2\pi} \pi q^{\frac{1}{3}}) \cos \varphi] \right] \: dx 
\right\} \: dt.
\label{e:chlang}
\end{eqnarray}
The equations of motion are obtained by varying the Lagrangian (\ref{e:chlang})
with respect to the parameters $\pi$, $\varphi$, $p$ and $q$.  These 
variational equations will consist of a partial differential equation for 
the field coupled to ordinary differential equations for the fluctuations. 
For the field the variational equation is
\begin{equation}
\frac{\partial^{2} \varphi}{\partial t^{2}} - \frac{\partial^{2}
\varphi}{\partial x^{2}} + \left( 1 - \frac{\Omega_{0}}{2} q^{\frac{1}{3}}\right) 
\sin {\varphi} = 0.
\label{e:psineg}
\end{equation}

The equations for $p$ and $q$ are derived from the Lagrangian
\begin{equation}
{\cal L} = \frac{2c_{1}L\Omega_{0}}{3\pi} \int _{0}^{T} \left[ \dot{q} p
- 3 H(p,q) \right] \: dt,
\label{e:pqlang}
\end{equation}
where the Hamiltonian for the fluctuations is given by 
\begin{equation}
H(p,q) = p^{2} q^{-5/3} + q^{1/3} \frac{1}{2L} \int _{-L}^{L} 
\cos \varphi \: dx - \frac{1}{8} q^{-1/3} + \frac{K}{8c_{1} \Omega^{2}_{0}} q^{-2/3}.
\label{e:hamfl}
\end{equation}
The equations of motion for $p$ and $q$ are then given by Hamilton's
equations as
\begin{eqnarray}
\dot{q} & = & \frac{\partial H}{\partial p} = 6p q^{-5/3} \label{e:peqn} \\
\dot{p} & = & -\frac{\partial H}{\partial q} = 5 p^{2} q^{-8/3}
              + \frac{2}{3} \beta q^{-5/3} - \frac{1}{24} q^{-4/3}
              - \frac{q^{-2/3}}{2L} \int_{-L}^{L} \cos \varphi \: dx,
\label{e:qeqn}
\end{eqnarray}
where
\begin{equation}
\beta = \frac{3K}{8c_{1} \Omega^{2}_{0}}.
\label{e:beta}
\end{equation}
Since $K$ is assumed to be small, we will take $\beta = 2.5$ in the numerical
calculations of the next section.  It must be noted that for initial
conditions which have $\varphi \to 2 \pi$ as $x \to -\infty$ and 
$\varphi \to 0$ as $x \to \infty$, or vice versa, 
\begin{equation}
\Lambda:=\lim_{L \to \infty} 
\frac{1}{2L} \int _{-L}^{L} \cos \varphi \: dx = 1.
\label{e:lim1}
\end{equation}
It is therefore apparent that the Equations (\ref{e:qeqn}) for the 
fluctuations decouple from the Equation (\ref{e:psineg}) for the field. 

It is interesting at this point to compare our equations (\ref{e:psineg}), (\ref{e:peqn}) and
(\ref{e:qeqn}) with the corresponding equations (4.5) of Ref. \cite{cooper}.
Observe that in that reference the equation for the field takes the form
\begin{equation}
\varphi_{tt}-\varphi_{xx}+\frac{\alpha_0}{\beta}
e^{-\frac{\beta^2}{2}G(x,x,t)}\sin \beta\varphi=0.
\label{e:mam}
\end{equation}
Taking $\beta=1$ and assuming $G(x,x,t)\ll 1$, which amounts to choosing the initial conditions in the form of small fluctuations, we have 
\begin{equation}
e^{-\frac{1}{2}G(x,x,t)}\approx 1-\frac{1}{2}G(x,x,t).
\end{equation}
Clearly, substituting this last expression into (\ref{e:mam}) recovers our equation (\ref{e:psineg}), so the respective field equations agree for the above mentioned initial conditions and value of $\beta$.
Now, as for the second equation (4.5) in \cite{cooper} note that this is an infinite system of partial
differential equations for the operator valued function $\hat{\Psi}$, while
in our formulation, because of the assumption of spatial homogeneity and
the functional form chosen for the trial Green function, the system
simplifies to a problem of ordinary differential equations. 
Another difference between our approach and that followed in \cite{cooper} is that the equations proposed there in order to arrive at 
numerical solutions are integrodifferential equations, as opposed to the simple variational approximation we propose for obtaining solutions. 
It must be also remarked that our assumed homogeneity
of the Green function is consistent with the low momentum limit we have chosen. In this low momentum limit, the fluctuations do not resolve the fine scale of the
field and, to leading order, the configuration is a homogeneous
background for the fluctuation. 
To conclude this section we also consider important to stress the fact that in previous published work the interest has been on static solutions
allowing for arbitrary momentum of arbitrarily large fluctuations.  This leads to a different renormalized version of the gap equation
\cite{eboli1,eboli2,cooper} and, for large coupling $\beta$,
to a loss of stability (phase transition).
Our approximation does not capture this region, since we have assumed from the onset small fluctuations and small momentum. However it must be noted that our procedure could be extended to handle large momenta by choosing different trial functions for $G$ and $\Sigma$, similar to the ones used in \cite{eboli2}. Also, due to the special form of the potential in the equations, the Gaussian integral may be evaluated to a better degree of approximation, thus allowing to include fluctuations of a larger amplitude. This program is currently under investigation and will be reported subsequently.

In the following section we undertake a detailed study of the dyamics described by the equations (\ref{e:psineg}), (\ref{e:peqn}) and (\ref{e:qeqn}).

\section{SOLUTIONS}

The system of the Sine-Gordon equation (\ref{e:psineg}) and the equations
(\ref{e:qeqn}) and (\ref{e:peqn}) for $q$ and $p$ describe the coupling 
of the field to the fluctuations and the corresponding feedback.  Note that 
the fluctuations have been assumed to be small. However, they are
allowed to feed back onto the basic field configuration. 
We shall now use these equations to describe in a nonperturbative manner the nonlinear evolution of some special field configurations.

\subsection{Quantum Stability of the Single Soliton}

We begin by studying the stability of the Soliton solution 
(\ref{e:sgsol}) 
under a class of initial values for $p$ and $q$.  Note that a small
value of $q$ represents a small variance.  Stability is then assured in the
model by the fact that $q$ remains small and that the field maintains its 
identity as a localized structure.

Numerical integrations of the Sine-Gordon equation (\ref{e:psineg}) and
the Equations (\ref{e:qeqn}) and (\ref{e:peqn}) for $p$ and $q$ have been 
performed for a wide range of initial conditions and typical behaviors 
are shown in Figures \ref{f:sol} and \ref{f:solmax}.  In Figure \ref{f:sol} 
a numerical solution for $\Phi = \varphi _{x}$ for the Skyrmion is shown and 
in Figure \ref{f:solmax} the behavior of $a$, the maximum  of $\Phi$, is shown.  
It can be seen that the fluctuations of the Skyrmion produce radiation, but 
that the field eventually stabilizes.  This can be clearly seen from the maximum 
behavior shown in Figure \ref{f:solmax}.  The stabilization onto a modulated small 
oscillation of the Skyrmion amplitude can be clearly seen.  These results 
exhibit the strong stability of the Skyrmion with respect to fluctuations.  
It is possible to understand this behavior by making use of the modulation 
theory given in \cite{minzoni} and \cite{whitham} by means of the following 
argument.  If the scales for $p$ and $q$ are slow, we may take as an 
approximate solution
\begin{equation}
\varphi = \varphi \left( (1 - \frac{\Omega_{0}}{2} q^{\frac{1}{3}})^{\frac{1}{2}}
x \right),
\label{e:appsoln}
\end{equation}
which satisfies 
\begin{equation}
\cos \varphi = 1 - \frac{1}{2} \dot{\varphi}^{2}.
\label{e:cosphi}
\end{equation}
Hence as $L \rightarrow \infty$ we have from (\ref{e:lim1}) that
$\Lambda=1$.
With this, the equations for $p$ and $q$ are derivable from the Hamiltonian
\begin{equation}
H(p,q)=3\left( p^{2} q^{-\frac{5}{3}} +
q^{1/3} - \frac{1}{8} q^{-1/3} + \frac{K}{4\Omega^{2}_{0}} \: q^{-2/3} \right).
\label{e:appham}
\end{equation}
Thus the orbits in the $(p,q)$ plane are just the level lines of
\begin{equation}
H(p,q) = E.
\label{e:orbits}
\end{equation}
The orbits of the $(p,q)$ system are then given by
\begin{eqnarray}
p^{2} & = & q^{5/3} \left[ E - \left( \frac{K}{4\Omega^{2}_{0}} 
\: q^{-2/3}
+ q^{1/3} - \frac{1}{8} q^{-1/3}\right) \right] \nonumber \\
 & = & q^{5/3} \left[ E - V(q) \right] . \label{e:pqham}
\end{eqnarray}
The potential $V(q)$ has a minimum which gives an oscillatory solution
for $p$ and $q$, so that the width of the Skyrmion and thus the amplitude
of $\Phi = \varphi _{x}$ oscillate in time.  The numerical solution shown
in Figure 1 shows that the radiation, which is not taken into account in
this approximation, stabilizes the oscillations onto a limit cycle.  

This strongly nonlinear mechanism accounts for the stability of the
Skyrmion.  In fact it is the feedback of the field on the fluctuations
which produces the term $q^{1/3}$ in $V(q)$ and it is this term which 
stabilizes the motion.  The potential $V(q)$ has a maximum for small $q$.
For energies $E$ larger than this maximum, the fluctuations $q$ increase and 
the field structure is destroyed.  However the value of $q$ for this to occur 
is too small to be consistent with the coherent state approximation.  The 
model is therefore self-consistent and provides an explanation of how 
nonlinear interactions are responsible for the quantum stability of the field.  

The approximate solution above does not take into account the radiation
produced by the oscillating Skyrmion and so this approximate solution
will not give the baryon settling onto a limit cycle solution.  To take account
of the radiation the ideas of Smyth and Worthy \cite{annette} can be used.
In this work the effect of shed dispersive radiation on the evolution of
a single pulse for the Sine-Gordon equation was treated.  To take account
of the radiation we proceed as in \cite{annette}, indicating only the main 
differences from this work.

The Lagrangian density for the Sine-Gordon equation is
\begin{equation}
L = \frac{1}{2} \varphi_{t}^{2} - \frac{1}{2}
\varphi _{x}^{2} - \left( 1 - \frac{\Omega_{0}}{2} q^{1/3} \right)
\left( 1 - \cos \varphi \right) .
\label{e:sglang}
\end{equation}
To obtain an approximate solution of the Sine-Gordon equation, the trial 
function
\begin{equation}
\varphi = -4 \arctan e^{-x/w(t)},
\label{e:trial2}
\end{equation}
which is a soliton-like pulse with varying width $w(t)$, is substituted into 
the averaged Lagrangian
\begin{eqnarray}
\bar{L} & = & \int_{-\infty}^{\infty} L \: dx \nonumber \\
        & = & \frac{\pi^{2}}{3} \frac{w'^{2}}{w} - \frac{4}{w} - 
4\left( 1 -  \frac{\Omega_{0}}{2} q^{1/3} \right) w ,
\label{e:annlang}
\end{eqnarray}
as in \cite{annette}.  In this approximation the Hamiltonian for $p$ and $q$ again does not change due to (\ref{e:lim1})

The effect of the radiation shed by the evolving soliton is determined by 
finding an appropriate solution of the linearized Sine-Gordon equation 
\cite{annette}.  The effect of this radiation is then to modify the 
Euler-Lagrange equation for $w(t)$.  It is noted from the numerical solution 
of Figure \ref{f:sol} that the radiation $\tilde{\varphi}$ is of small 
amplitude compared with the soliton.  Therefore following \cite{annette} 
we consider the linearized Sine-Gordon equation
\begin{equation}
\frac{\partial ^{2} \tilde{\varphi}}{\partial t^{2}} - 
\frac{\partial ^{2} \tilde{\varphi}}{\partial x^{2}} + \left( 1 - \frac{\Omega_{0}}{2} q^{1/3}(t) \right) \tilde{\varphi} = 0 
\label{e:linsg}
\end{equation}
for the radiation $\tilde{\varphi}$.  This equation is solved together with 
appropriate source conditions at the pulse at $x=0$.  Since $\Omega_{0} \ll 1$,
\begin{equation}
\frac{d}{dt} \left( 1  - \frac{\Omega_{0}}{2} q^{1/3}(t) \right) \ll 1.
\label{e:ll1}
\end{equation}
It is then possible to obtain an expression for the radiation by making the 
adiabatic approximation that $1 - \frac{\Omega_{0}}{2} q^{1/3}$ is constant to 
leading order.  The effect of the radiation can then be found from the 
expression of \cite{annette} by a suitable re-scaling.  In this manner we 
find that the equations governing the evolution of the soliton, including 
the effect of radiation, are
\begin{eqnarray}
\frac{2\pi^{2}}{3w} \frac{d^{2}w}{dt^{2}} - \frac{\pi^{2}}{3w^{2}} \left( 
\frac{dw}{dt} \right) ^{2} - \frac{4}{w^{2}} & & \nonumber \\
\mbox{} + 4 \left( 1 - \frac{\Omega_{0}}{2} q^{1/3} \right) & = &
\frac{1}{\sqrt{\lambda}} \left[ -\frac{2\pi^{2}}{3w\sqrt{\lambda} \: t} 
\frac{dw}{dt} \right. \nonumber \\
 & & \mbox{} + \left. \frac{2\pi^{2}}{3\sqrt{wt}} \int_{0}^{t}
J_{1}(\sqrt{\lambda}(t-\tau)) \frac{w'(\tau)}{\sqrt{\tau w(\tau)}} \: d\tau 
\right] \nonumber \\
\frac{dq}{dt} & = & 6 p q^{-5/3} \label{e:radeqn} \\
\frac{dp}{dt} & = & 5 p^{2} q^{-8/3} + \frac{2}{3} \beta q^{-5/3} - q^{-2/3} 
+ \frac{1}{9} q^{-4/3}, \nonumber
\end{eqnarray}
where 
\begin{equation}
\lambda = 1 - \frac{\Omega_{0}}{2} q^{1/3}.
\label{e:lam}
\end{equation}
These equations were integrated numerically.
 Comparisons between solutions of
these equations and the full numerical solution of the Sine-Gordon equation 
for the amplitude $a$ of $\Phi = \varphi _{x}$ and
$q(t)$ for the fluctuations 
are shown in Figure \ref{f:ampcomp}.
It can be seen that the amplitude agreement 
shown in Figure \ref{f:ampcomp} is good considering the
assumptions that were made 
to incorporate the radiation loss in the approximate equations.
It can be seen 
that the approximate equations provide a suitable
approximate solution for the 
full field behaviour using a finite dimensional approximation which includes 
radiation.  It is noted that, since $q(t)$ is periodic, the Sine-Gordon
Equation (\ref{e:psineg}) is subject to a parametric excitation.  However
the nonlinearity and radiation loss provide the necessary damping to enable
a limit cycle to be achieved.

\subsection{The collision of a wave with a static soliton}

As a final example we consider the scattering of a wave packet representing 
a pion with momentum $k$ with a static soliton, representing
a baryon originally at rest.  The problem is 
solved by numerically integrating the Sine-Gordon equation (\ref{e:psineg})
using the initial condition
\begin{eqnarray}
\varphi(x) & = & -4\arctan e^{-x}+f(x) \label{e:icf} \\
\frac{\partial\varphi}{\partial t} & = & g(x),
\label{e:icv}
\end{eqnarray}
where the functions $f$ and $g$ are given by
\begin{eqnarray}
f(x) & = & a\sin k(x+x_0),\quad |x+x_0|\leq\delta \label{e:f} \\
g(x) & = & -a\sqrt{k^2+1}\cos k(x+x_0),\quad |x+x_0|\leq\delta . \label{e:g}
\end{eqnarray}
This initial condition represents an incoming meson with momentum $k$ 
impinging on a nucleon located at $x=0$.  A numerical solution for the 
scattering of the pion wavepacket can be seen in Figures \ref{f:solrad} to 
\ref{f:raddet}.  The initial condition (at $t=0$) is shown by the solid line 
in Figure \ref{f:solrad}.  In this Figure a reflected wave packet, a reorganized 
field configuration and a new packet shed by the baryon as a result of the 
interaction can be seen.  In Figures \ref{f:radmax} and \ref{f:raddet} the 
complicated evolution of the baryon amplitude is displayed.  This amplitude 
behavior is due to the interaction of the baryon with the packet.  The scattering 
then involves a reorganization of the field, which is not taken into account when 
the particles are taken to be point particles.  The description of the 
interaction of the baryon with radiation using a multi-phase solution of 
the Sine-Gordon equation is under investigation at present.

\subsection{Collision of two solitons in the presence of a fluctuation}

Since the classical field equation is completely integrable, solitons interact 
elastically and do not change configuration.  The effect of quantum 
fluctuations on the collisions of solitons 
and this clean interaction will 
now be studied.

Figure \ref{f:two} shows the collision of two Solitons with equal and opposite
velocity.  The initial condition used was
\begin{equation}
\varphi = 2 \pi - 4 \arctan e^{-(x+x_{0}-vt)/\sqrt{1-v^{2}}} 
- 4 \arctan e^{-(x-x_{0}+vt)/\sqrt{1-v^{2}}}
\label{e:twosol}
\end{equation}
as $t \to -\infty$.  Since there is no classical solution with twice the 
baryon number and zero velocity, the effect of the quantum 
fluctuations is 
just to slightly modify the classical interaction.  
The Solitons again settle
down to a limit cycle for which the parametric resonance is balanced by
the radiation damping.

Figure \ref{f:anti} shows the collision of a Soliton
 and an anti-soliton.
The initial condition is
\begin{equation}
\varphi = -4 \arctan \left[ \frac{\alpha}{\sqrt{\alpha ^{2} -1}}
          \frac{\sinh \sqrt{\alpha^{2}-1} \: t}{\cosh \alpha x} \right].
\label{e:anti}
\end{equation}
Again this interaction does not produce disintegration, just a modification
of the classical interaction.  

Finally the susceptibility to disintegration of the breather-type
configuration 
\begin{equation}
\varphi = -4 \arctan \left[ \frac{\alpha}{\cosh \alpha x} 
\frac{\sin \sqrt{1-\alpha^{2}} \: t}{\sqrt{1-\alpha^{2}}} \right]
\label{e:breath}
\end{equation}
with frequency $\sqrt{1-\alpha^{2}}$ is studied.  From the numerical solution
shown in Figure \ref{f:breath} it can be seen that the breather is stable
with respect to quantum fluctuations. 

The solutions displayed in Figure 4 show that the reduction of the Skyrme model 
to the Sine-Gordon equation is too severe for treating collisions.  In order 
to obtain non-trivial collision and fusion processes, such as those possible 
for the nonlinear Schr\"odinger equation, reductions of the Skyrme model 
which retain more internal degrees of freedom must be derived.  

\section{CONCLUSIONS AND SUGGESTIONS FOR FURTHER RESEARCH}

We have formulated the quantum field problem for the Sine-Gordon
equation which is related to the (dimensionally reduced) Skyrme model.
Using the coherent state approximation for the solution of the
functional Schr\"odinger equation, we obtain a solution of the partial
differential equation for the (quantum corrected, semi-classical)
field, which is coupled to ordinary differential equations for the 
fluctuations.  Other quantizations for nonlinear fields keep only finitely 
many degrees of freedom (minisuperspace approximation), which are then 
quantized in a canonical way.  

The first problem considered in the present work was the stability with
respect to quantum fluctuations of a Soliton.  Both numerical and asymptotic 
solutions were considered.  It was found that the nonlinear saturation of the 
field equation together with the loss of radiation balanced the parametric 
excitation of the fluctuations.  The fluctuations in turn were controlled 
by the shape of the field.  The good agreement found between numerical
and asymptotic solutions suggests that finite dimensional approximations
to the dynamics of the Skyrmion model, such as those used in \cite{gisiger},
are also good approximations to the full dynamics of more complicated 
problems, such as those treated in \cite{gisiger}.  

The scattering of a wave by a static soliton was also studied.
  The numerical
results obtained show well defined waves and a Skyrmion after collision,
which suggests the possibility of using multi-phase solutions, such as
those of \cite{15}, to understand this scattering process.  

Finally several collision processes were studied.  It was found that the
reduced Skyrme model cannot account for the collision and fusion of baryons.
Therefore the study of the fusion of Skyrmions into a toroidal configuration
requires a uniform solution which interpolates between the torus and the
individual Skyrmions.  The possibility of using the solutions given in
\cite{13,14} is currently under study.
It must be noted that more sophisticated numerical formulations such as
the ones proposed in \cite{cooper} must produce, in the limit of low
momentum, solutions comparable to our results.

To conclude, we note that the techniques described in this work can be applied
to the study of low dimensional black holes.  Indeed, an old observation that 
Sine-Gordon theory and 2-dimensional spaces of constant curvature are very 
closely related has recently found an interesting application to gravity in
$1+1$ dimensions.  More precisely, Gegenberg and Kunstatter \cite{gegen} 
have noticed that when a two dimensional Lorenzian metric is parameterized as
\begin{equation}
ds^2 = -\sin^2 (u/2) dt^2 + \cos^2 (u/2) dx^2,
\label{e:metric}
\end{equation}
then the condition of constant curvature is equivalent to the condition
that $u$ satisfies the {\it Euclidean} Sine-Gordon equation.  On the other
hand, the so-called Jackiw-Teitelboim theory in two dimensions 
\begin{equation}
I = \int \phi(R - \Lambda )\sqrt{-g} \: dt dx.
\label{e:jack}
\end{equation}
has as solutions space-times of constant curvature $R=\Lambda$.  Furthermore,
the one-soliton solution of the Sine-Gordon equation has been found to 
represent (a patch of) a black hole solution of the (Jackiw-Teitelboim) theory
\cite{gegen}.  That a constant curvature space-time can be interpreted as 
a black hole is not unique to two dimensions.  The $1+1$ Jackiw-Teitelboim 
black hole can indeed be interpreted as a dimensionally reduced BTZ
(non-rotating) black hole and many of its properties (including 
thermodynamics) have been studied \cite{lemos}.

To perform an analysis similar to the one presented in the present work
for the Euclidean Sine-Gordon equation is cumbersome, since the
equation is now elliptic and does not accept a well-posed initial value
formulation.  However it is possible to work in the framework of a well-posed 
problem if one chooses a different parametrization for the
two dimensional space-time as follows
\begin{equation}
ds^2 = -\sinh^2 (u/2) dt^2 + \cosh^2 (u/2) dx^2.
\label{e:dssinh}
\end{equation}
In this case, the constant curvature condition reduces to the 
{\it Lorenzian} Sinh-Gordon equation.  It is then possible to analyze 
the quantum stability of a black hole solution using the functional methods 
presented in this article.  This work will be reported elsewhere.

As a final remark we point out that the quantum equations for a classical 
field obtained using the functional Schr\"odinger equation and the 
coherent state approximation will always have the same structure.  Namely
the classical equations for the field with renormalized (fluctuating) 
parameters and equations for the (parameters of the) fluctuations which 
are non-local in the fields will always be obtained.

\section{ACKNOWLEDGEMENTS}

This work was supported in part by UNAM DGAPA Project No.\ IN 106097.
  A.C. was also supported by Conacyt Proyect No.\ I25655-E.


\newpage

\section*{Figure Captions}

\renewcommand{\theenumi}{Figure \hspace{1em}\arabic{enumi}}

\begin{enumerate}

\item  Stability of a single Soliton to fluctuations.  Solution of Sine-Gordon
       equation (\ref{e:psineg}) and equations
 (\ref{e:qeqn}) for $p$ and $q$ with 
       $\Omega_{0}=0.6$ and $\beta=2.5$.  The initial conditions are 
       $q=1.0$ and $p=0.0$ and $v=0$ in the soliton solution 
(\ref{e:sgsol}) at 
       $t=0$.

       \begin{enumerate}

       \item  Soliton.  ---: initial condition; ~--~--~--~: Soliton at 
              $t=100$.

       \item  Evolution of maximum $a$ of $\Phi = \varphi _{x}$.

       \end{enumerate}

\item  Stability of a single Soliton to fluctuations.
  Comparison between the 
       full numerical solution of the Sine-Gordon equation 
(\ref{e:psineg}) and 
       equations (\ref{e:qeqn}) for $p$ and $q$ and the approximate Equations        (\ref{e:radeqn}).  
$\Omega_{0}=0.6$, $\beta=2.5$.  The initial conditions 
       are $q=1.0$, $p=0.0$ and $v=0$ at $t=0$.  Amplitude $a$ of 
       $\Phi = \varphi_{x}$.  Full numerical solution: ---; approximate 
       solution: ~--~--~--~.

\item  Scattering of a wave packet (pion) with a baryon.  Solution of 
       Sine-Gordon equation (\ref{e:psineg}) and equations (\ref{e:qeqn}) 
       for $p$ and $q$ with $\Omega_{0}=0.6$ and $\beta =2.5$.  The initial 
       conditions are given by (\ref{e:icf}), (\ref{e:icv}) and $a=0.1$, 
       $k=1.0$, $\delta =4.0$ and $x_{0}=20$ in (\ref{e:f}) and (\ref{e:g}).  
       Also $q=1.0$ and $p=0.0$ at $t=0$.

       \begin{enumerate}

       \item  Solution at $t=50$.

       \item  Evolution of maximum $a$ of $\Phi = \varphi _{x}$.

       \item  Detail of evolution of maximum $a$ of 
              $\Phi = \varphi _{x}$ from
              $t=15$ to $t=40$.

       \end{enumerate}

\item  Collisions of Solitons.  
Initial conditions have $q=1.0$ and $p=0.0$.
       $\Omega_{0}=0.6$.

       \begin{enumerate}

       \item  Two Solitons.  Initial condition (\ref{e:twosol}) with
              $x_{0}=15$ and $v=0.2$.  Initial condition ($t=0$): ---;  
              solution at $t=150$: ~--~--~--~--~.

       \item  Soliton and an anti-Soliton.  Initial condition (\ref{e:anti})
              with $\alpha = 1.2$.  Initial condition ($t=-15$): ---;
              solution at $t=15$: ~--~--~--~--~.

       \item  Bound state of a Soliton and an anti-Soliton.  Initial
              condition (\ref{e:breath}) with $\alpha = 0.98$.  Initial 
              condition ($t=-5$): ---; solution at $t=45$: ~--~--~--~--~.

       \end{enumerate}

\end{enumerate}


\newcounter{figalp}
\setcounter{figalp}{1}
\renewcommand{\thefigure}{\arabic{figure}(\alph{figalp})}

\vspace*{\fill}
\begin{figure}[t]
\epsfxsize=0.8\textwidth
\centerline{\epsfbox{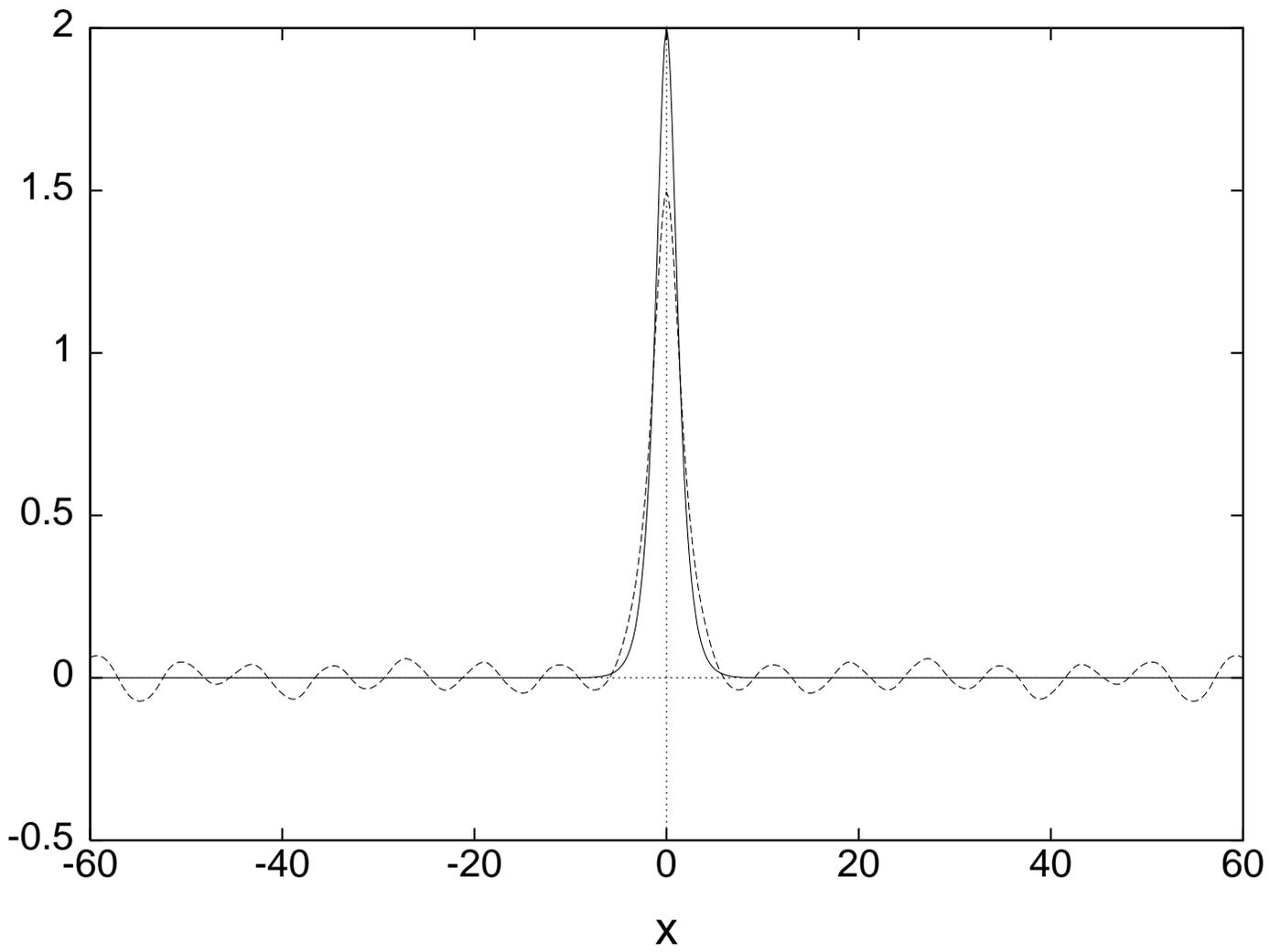}}
\caption{ --
           ``The Effect of Low  Momentum \ldots'' by G. Cruz et.\ al.}
\label{f:sol}
\end{figure}
\addtocounter{figure}{-1}
\addtocounter{figalp}{1}
\vspace*{\fill}



\vspace*{\fill}
\begin{figure}[t]
\epsfxsize=0.8\textwidth
\centerline{\epsfbox{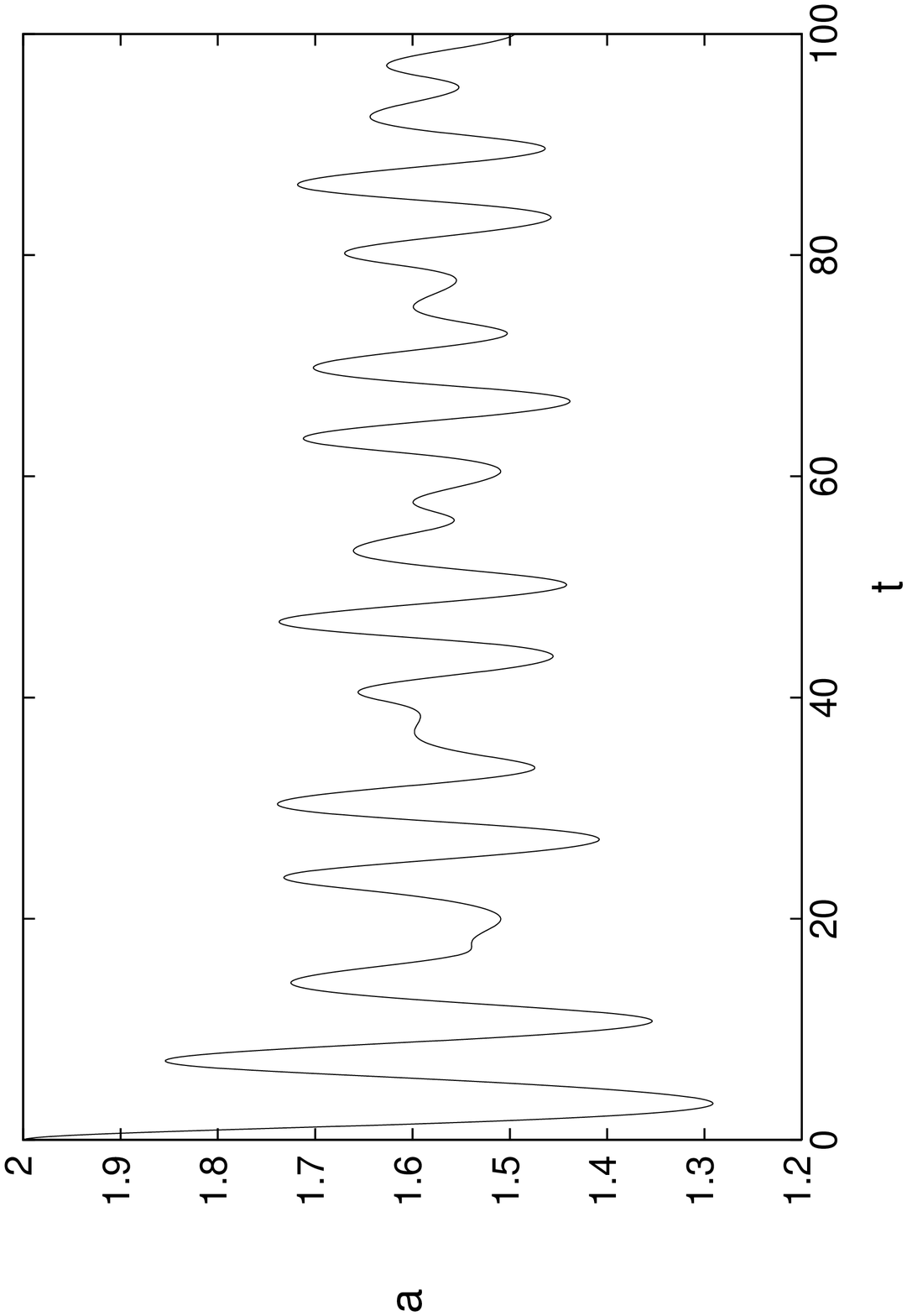}}
\caption{ --
           ``The Effect of Low  Momentum \ldots'' by G. Cruz et.\ al.}
\label{f:solmax}
\end{figure}
\vspace*{\fill}
\addtocounter{figalp}{-1}


\renewcommand{\thefigure}{\arabic{figure}}

\vspace*{\fill}
\begin{figure}[t]
\epsfxsize=0.8\textwidth
\centerline{\epsfbox{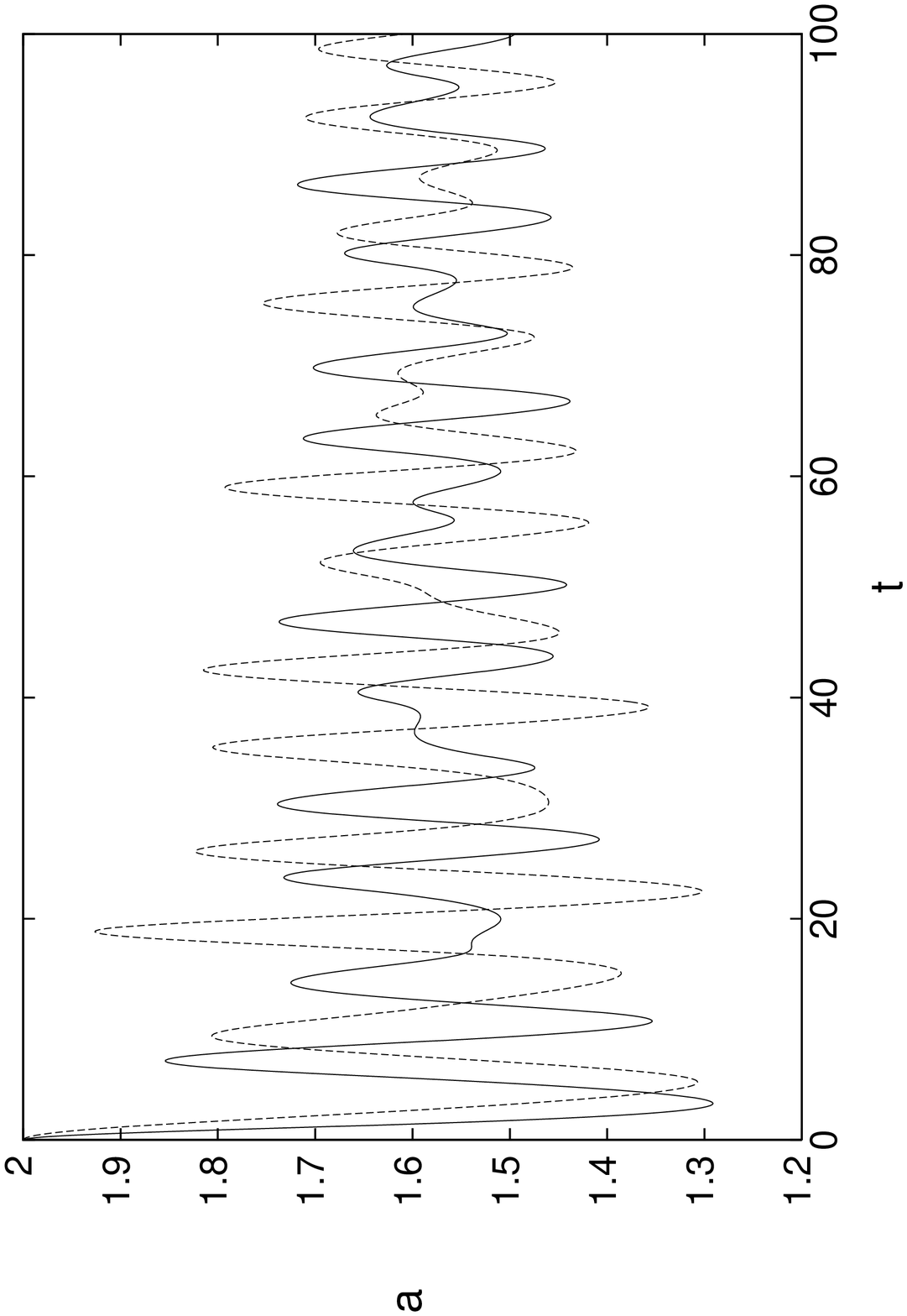}}
\caption{ --
           ``The Effect of Low  Momentum \ldots'' by G. Cruz et.\ al.}
\label{f:ampcomp}
\end{figure}
\vspace*{\fill}


\renewcommand{\thefigure}{\arabic{figure}(\alph{figalp})}

\vspace*{\fill}
\begin{figure}[t]
\epsfxsize=0.8\textwidth
\centerline{\epsfbox{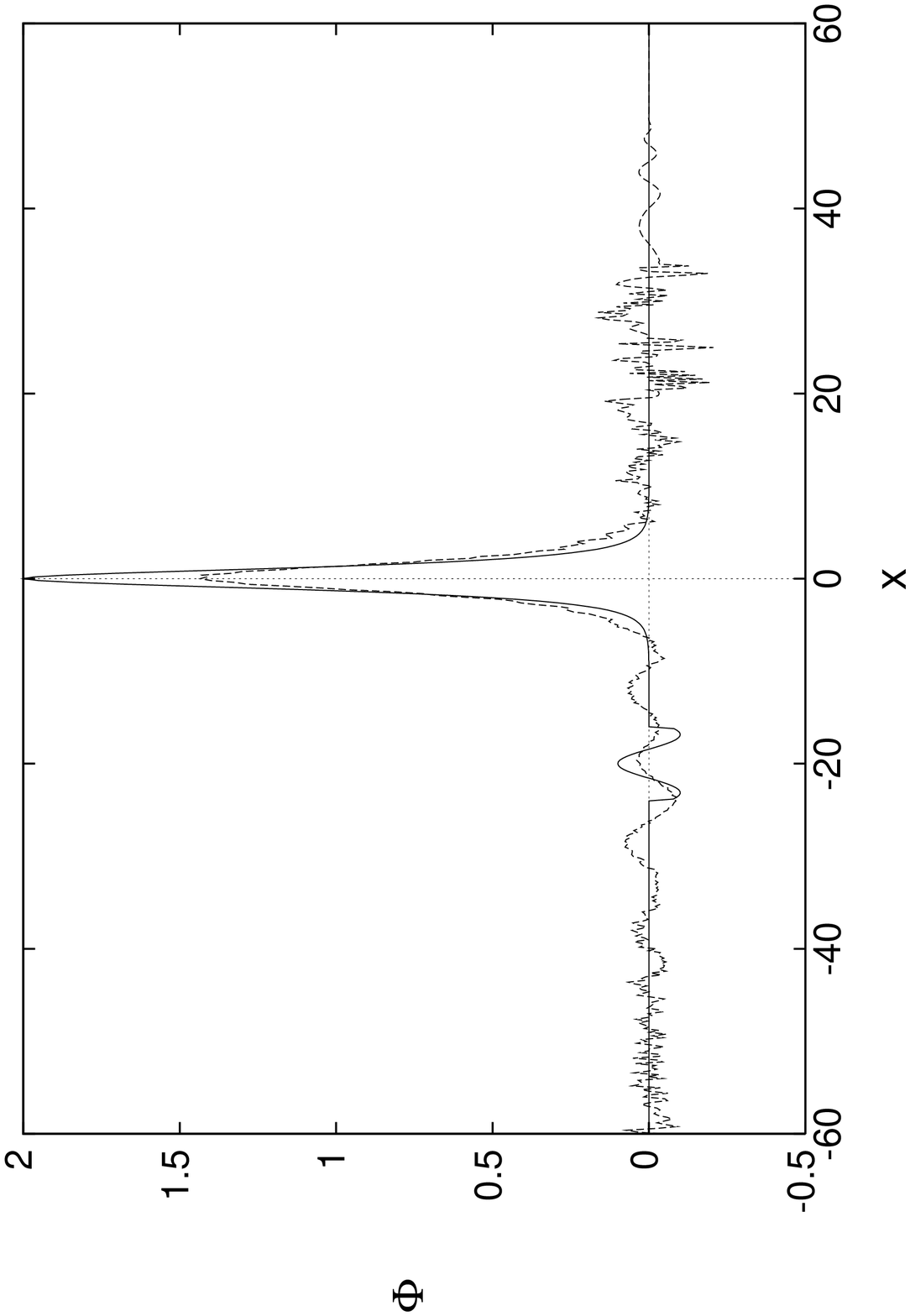}}
\caption{ --
           ``The Effect of Low  Momentum \ldots'' by G. Cruz et.\ al.}
\label{f:solrad}
\end{figure}
\vspace*{\fill}
\addtocounter{figure}{-1}
\addtocounter{figalp}{1}



\vspace*{\fill}
\begin{figure}[t]
\epsfxsize=0.8\textwidth
\centerline{\epsfbox{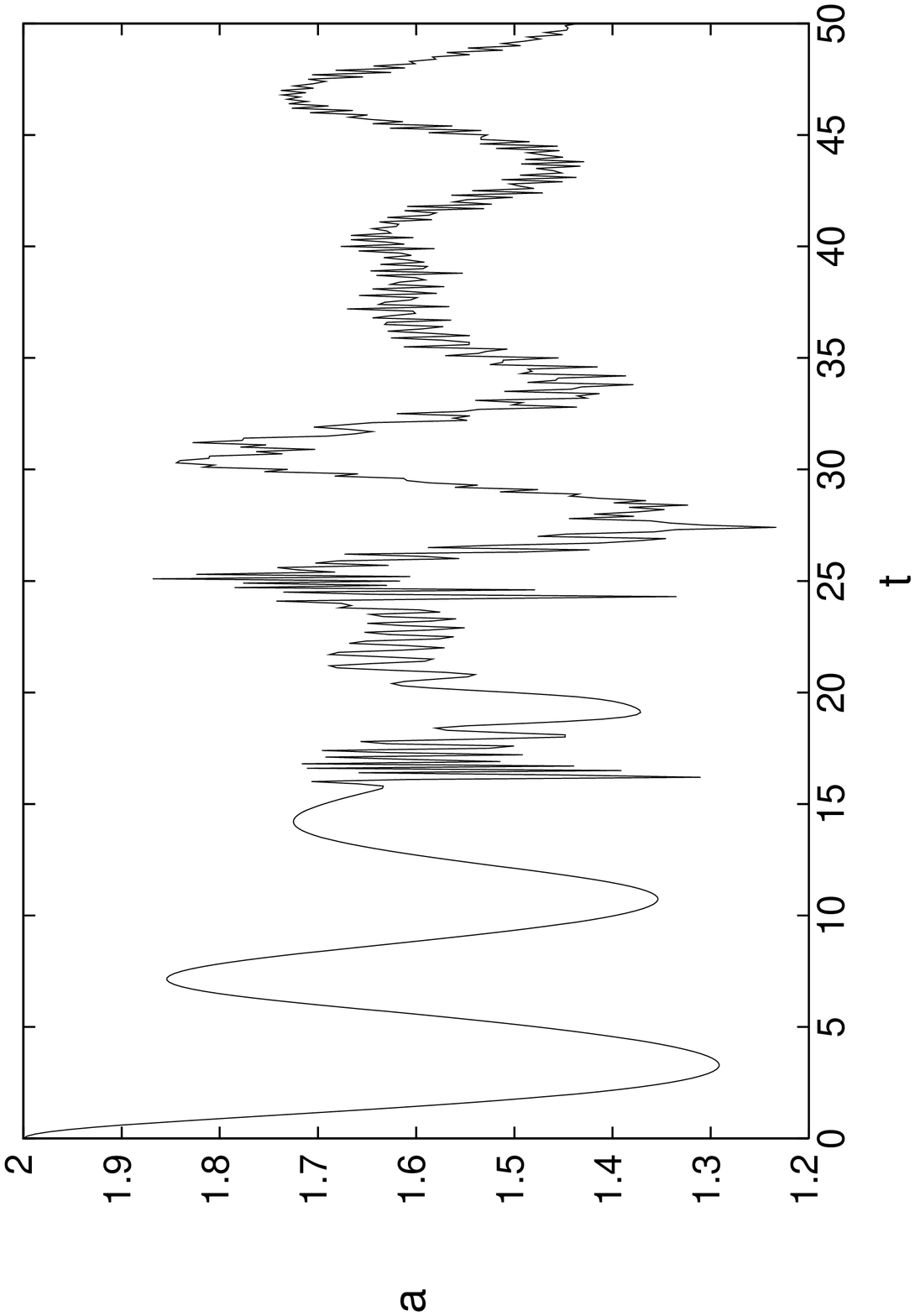}}
\caption{ --
           ``The Effect of Low  Momentum \ldots'' by G. Cruz et.\ al.}
\label{f:radmax}
\end{figure}
\vspace*{\fill}
\addtocounter{figure}{-1}
\addtocounter{figalp}{1}



\vspace*{\fill}
\begin{figure}[t]
\epsfxsize=0.8\textwidth
\centerline{\epsfbox{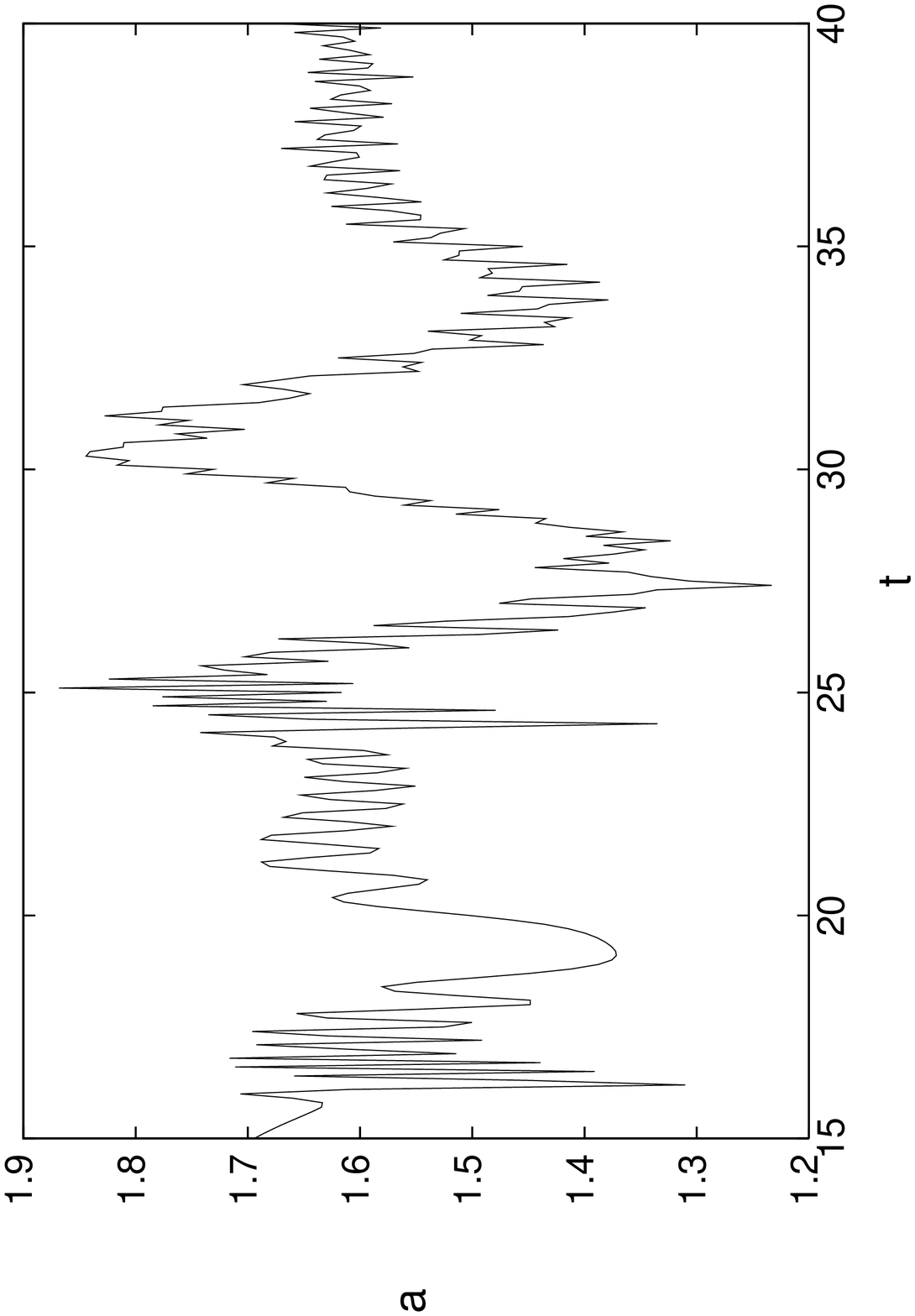}}
\caption{ --
           ``The Effect of Low  Momentum \ldots'' by G. Cruz et.\ al.}
\label{f:raddet}
\end{figure}
\vspace*{\fill}
\addtocounter{figalp}{-2}



\vspace*{\fill}
\begin{figure}[t]
\epsfxsize=0.8\textwidth
\centerline{\epsfbox{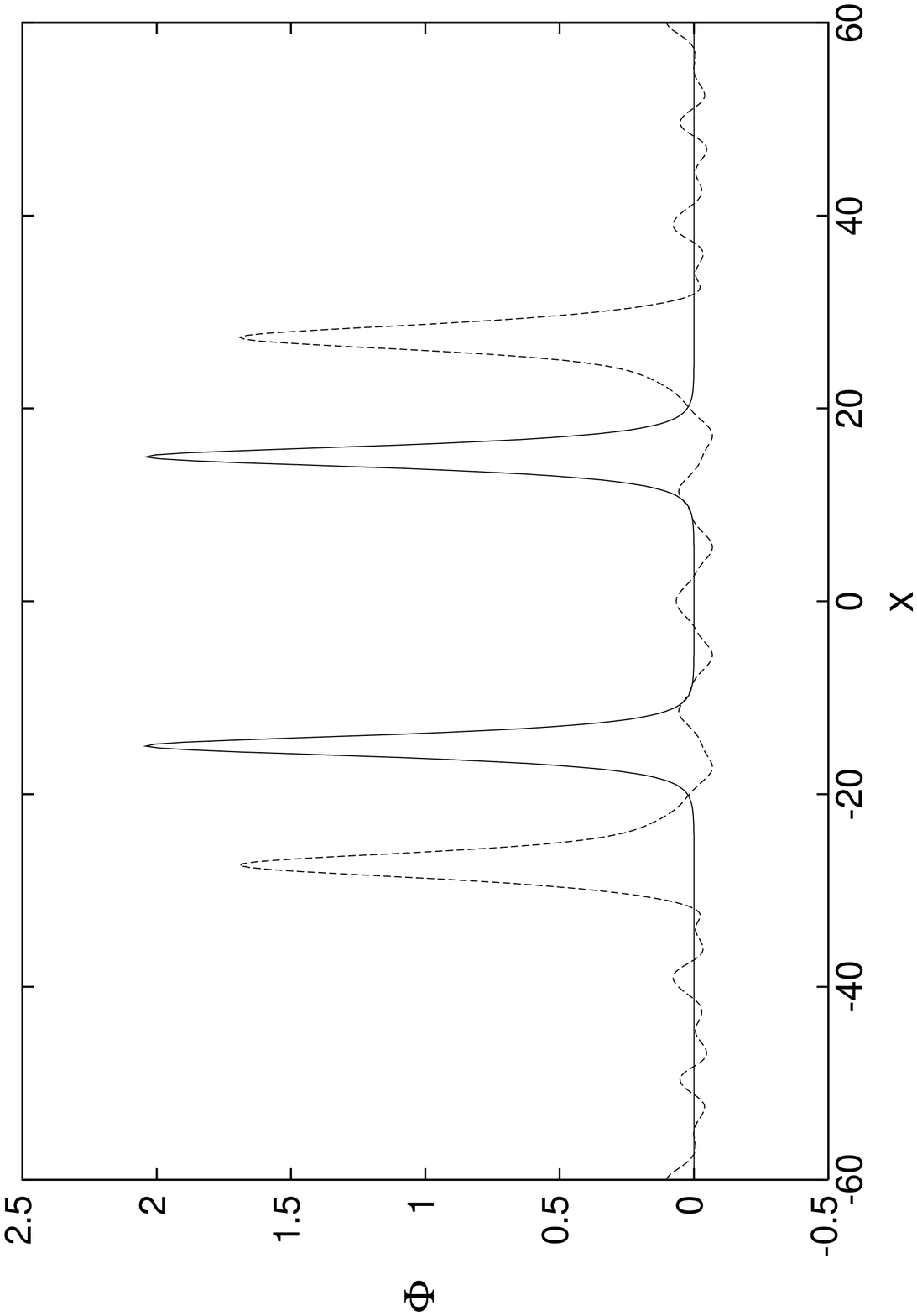}}
\caption{ --
           ``The Effect of Low  Momentum \ldots'' by G. Cruz et.\ al.}
\label{f:two}
\end{figure}
\vspace*{\fill}
\addtocounter{figure}{-1}
\addtocounter{figalp}{1}



\vspace*{\fill}
\begin{figure}[t]
\epsfxsize=0.8\textwidth
\centerline{\epsfbox{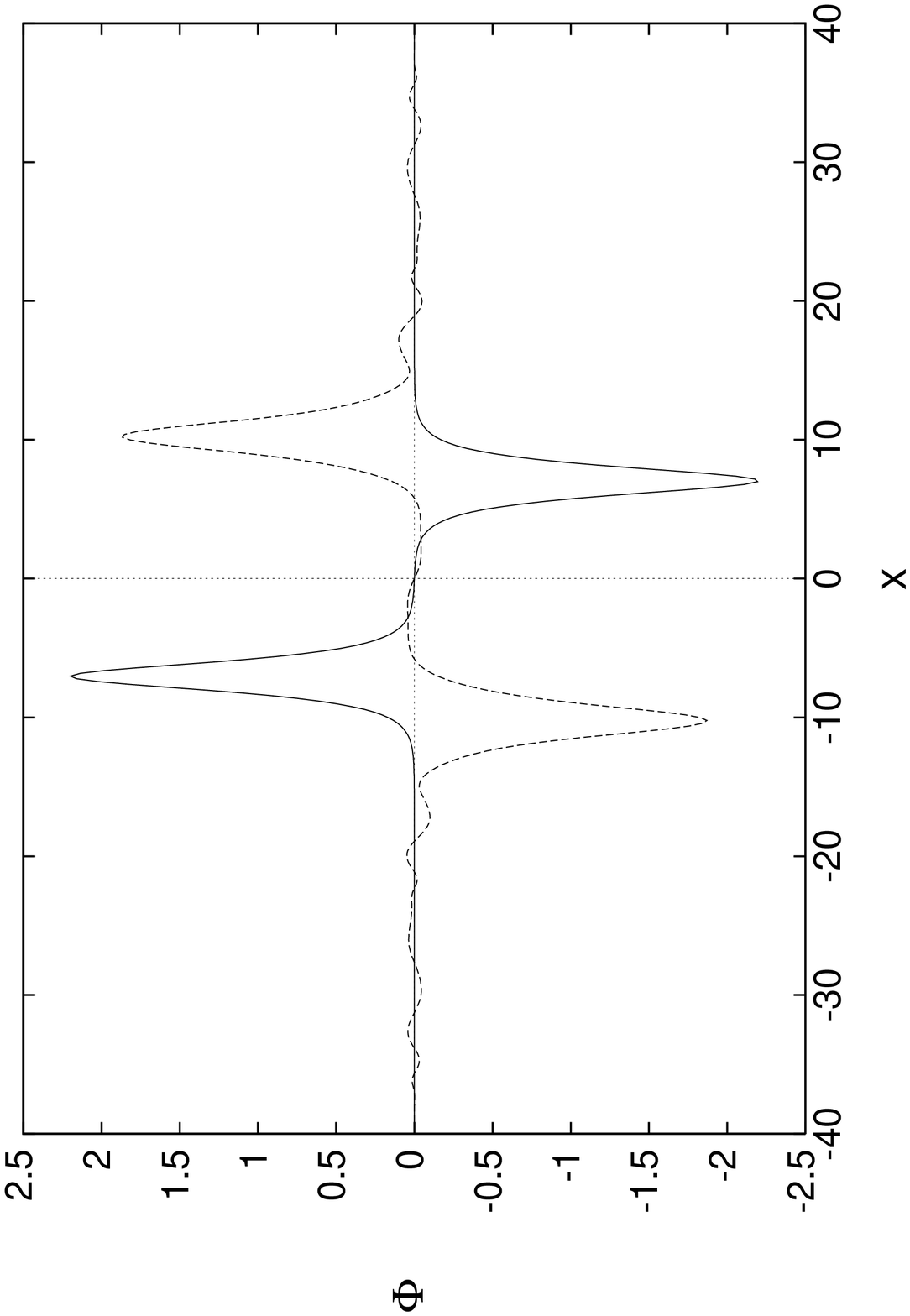}}
\caption{ --
           ``The Effect of Low  Momentum \ldots'' by G. Cruz et.\ al.}
\label{f:anti}
\end{figure}
\vspace*{\fill}
\addtocounter{figure}{-1}
\addtocounter{figalp}{1}



\vspace*{\fill}
\begin{figure}[t]
\epsfxsize=0.8\textwidth
\centerline{\epsfbox{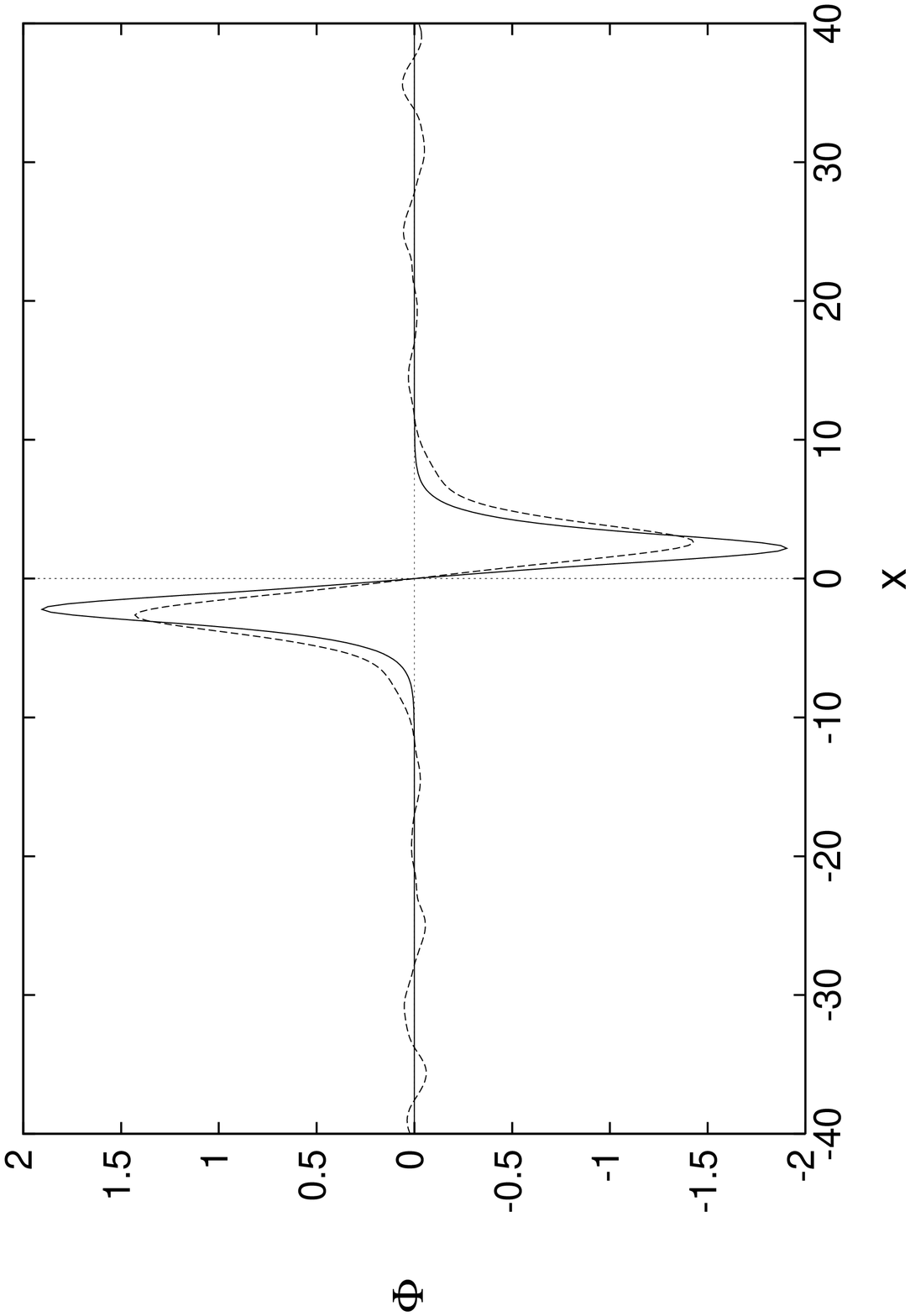}}
\caption{ --
           ``The Effect of Low  Momentum \ldots'' by G. Cruz et.\ al.}
\label{f:breath}
\end{figure}
\vspace*{\fill}
\addtocounter{figalp}{-2}

\end{document}